\documentclass[aps,prd,twocolumn,showpacs,eqsecnum,A4,amsmath,nofootinbib]{revtex4-1}

\usepackage{graphicx}

\def \d {{\rm d}}

\begin{document}

\title{\bf The Linet--Tian solution with a positive cosmological constant\\ in four and higher dimensions}

\author{J. B. Griffiths}
\email{J.B.Griffiths@lboro.ac.uk}
\affiliation{
  Department of Mathematical Sciences, Loughborough University, Loughborough,  Leics. LE11 3TU, U.K.}
\author{J. Podolsk\'y}
\email{podolsky@mbox.troja.mff.cuni.cz}
\affiliation{
  Institute of Theoretical Physics,  Faculty of Mathematics and Physics, Charles University in Prague,\\
  V Hole\v{s}ovi\v{c}k\'{a}ch 2, 180 00 Prague 8, Czech Republic}

\date{\today}

\begin{abstract}
\noindent
The static, apparently cylindrically symmetric vacuum solution of Linet and Tian for the case of a positive cosmological constant~$\Lambda$ is shown to have toroidal symmetry and, besides~$\Lambda$, to include three arbitrary parameters. It possesses two curvature singularities, of which one can be removed by matching it across a toroidal surface to a corresponding region of the dust-filled Einstein static universe. In four dimensions, this clarifies the geometrical properties, the coordinate ranges and the meaning of the parameters in this solution. Some other properties and limiting cases of this space-time are described. Its generalisation to any higher number of dimensions is also explicitly given.
\end{abstract}


\maketitle

\section{Introduction}

The well-known Levi-Civita solution \cite{LeviC19} describes the vacuum field exterior to an infinite cylinder of matter. In its general form, it contains both a parameter $\sigma$ which, for values in the range ${(0,{1\over4})}$, may be interpreted as the mass per unit length of the source, and also a conicity parameter. A generalisation of this to include a non-zero cosmological constant~${\Lambda}$, which can be either positive or negative, was obtained by Linet~\cite{Linet86} and Tian~\cite{Tian86}. This metric is algebraically general and, as expected, locally approaches the Levi-Civita solution either as ${\Lambda\to0}$ or near the axis as ${\rho\to0}$. The Linet--Tian solution and its non-vacuum generalisations have been used to describe cosmic strings (see e.g.~\cite{Tian86}--\cite{BhaLah08}). It has also recently been extended to higher-dimensions \cite{SarTek09} for ${\Lambda<0}$. It therefore seems appropriate to analyse it in greater detail. Here, we will describe some of its basic properties --- its geometry, the range of its coordinates, the number of its independent parameters and its limits.

One might initially expect that, for ${\Lambda\ne0}$, the Linet--Tian solution would reduce to a de~Sitter or anti-de~Sitter background space as ${\sigma\to0}$ (see~\cite{Bonnor08} for ${\Lambda<0}$). However, this is not possible since the conformally flat de~Sitter and anti-de~Sitter spaces are not compatible with {\em static} cylindrical symmetry. (When expressed in cylindrical coordinates, the metrics for these space-times are explicitly time-dependent.) In fact, as shown by da~Silva {\it et al}.~\cite{dSWPS00}, this particular limit is a type~D space-time. It is a special member of the Pleba\'nski--Demia\'nski family with non-expanding repeated principal null directions, and hence also belongs to Kundt's class (see e.g.~\cite{GriPod09}).

As for the Levi-Civita solution, a physical interpretation of the Linet--Tian solution depends critically on matching it to suitable interior metrics which remove their curvature singularities. However, to date, the only known sources of this solution are cylindrical shells of Bi\v{c}\'ak and \v{Z}ofka~\cite{ZofBic08}.

The purpose of the present paper is to exhibit some further interesting properties of this solution for the case with ${\Lambda>0}$. This will be achieved by examining its invariance properties and by matching it across a toroidal hypersurface to a suitable region of the Einstein static dust-filled universe.

Because of its importance in this context, the Einstein static universe is first analysed geometrically and an appropriate cylindrical-type coordinate representation is introduced. The Linet--Tian solution with a positive cosmological constant is then investigated and matched to a toroidal section of the Einstein universe. Further properties and limiting cases are also described, including the extension of the metric to an arbitrary higher number $D$ of dimensions.

\section{The Einstein static universe}

The obvious candidate for an interior solution of a static cylindrically symmetric space-time with a positive cosmological constant is a cylindrical section of the Einstein static universe. This has a constant non-zero mass density~$\mu$, such that ${4\pi\mu=\Lambda}$, and zero pressure. However, since such a space-time is closed, an apparently cylindrical section is, in fact, toroidal.

The metric for the Einstein static universe is often given in global coordinates in the form
 \begin{equation}
 \d s^2=-\d t^2+a^2\Big(\d\chi^2+\sin^2\chi\,(\d\theta^2+\sin^2\theta\,\d\phi^2) \Big),
 \label{Estatmetric}
 \end{equation}
 where ${t\in(-\infty,\infty)}$, ${\chi\in[0,\pi]}$, ${\theta\in[0,\pi]}$ and ${\phi\in[0,2\pi)}$. This space-time is spatially closed and the constant~$a$, which represents the radius of constant-curvature 3-spheres, determines both the density $\mu$ and the value of the cosmological constant as
 $$ 4\pi\mu= \Lambda =\frac{1}{a^{\,2}}. $$

In fact, any Friedmann--Lema\^{\i}tre--Robertson--Walker universe can be expressed in terms of cylindrical or toroidal coordinates. (For a detailed description of this, see~\cite{BicGri96}.) For this particular case, the transformation
 $$ \hat\rho=\sin\chi\sin\theta, \qquad \tan\psi=\tan\chi\cos\theta, $$
 takes the metric (\ref{Estatmetric}) to the apparently cylindrical form\footnote{
 It has orthogonal spacelike Killing vectors $\partial_\phi$ and $\partial_{\psi}$ and a regular axis at ${\hat\rho=0}$. Near the axis it resemble flat space in cylindrical coordinates.}
 \begin{equation}
 \d s^2=-\d t^2+a^2\bigg( {\d\hat\rho^2\over1-\hat\rho^2}
 +(1-\hat\rho^2)\d\psi^2
 +\hat\rho^2\,\d\phi^2\bigg) ,
\label{Estatmetric3}
 \end{equation}
 where ${\hat\rho\in[0,1]}$ and ${\psi\in[0,2\pi)}$. Putting ${\hat\rho=\sin(\rho/a)}$ and ${\psi=z/a}$, the metric becomes
\begin{equation}
\d s^2=-\d t^2+ \d\rho^2 +\cos^2(\sqrt\Lambda\,\rho)\,\d z^2
 +{1\over\Lambda}\sin^2(\sqrt\Lambda\,\rho)\, \d\phi^2 ,
\label{Estatmetric5}
 \end{equation}
 in which $\rho$ can be seen to be a proper (cylindrical) radial distance. It may also be noted that this metric clearly approaches that of empty Minkowski space as ${\Lambda\to0}$.

\begin{figure}[ht]
\begin{center}
\includegraphics[scale=0.7, trim=5 5 5 5]{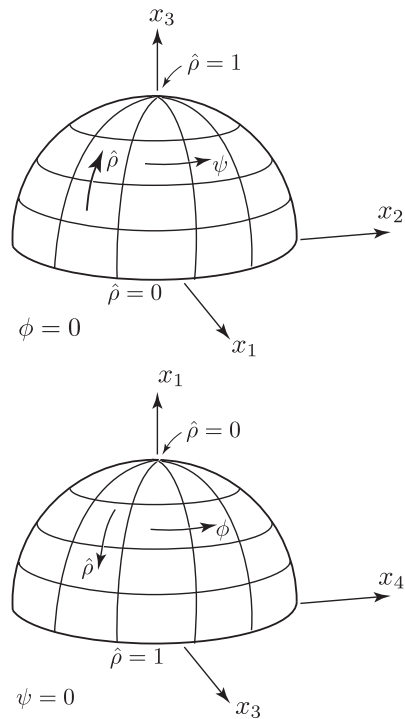}
\end{center}
\caption{\small The three-sphere in a four-dimensional Euclidean space is here represented by two sections: those on which ${\phi=0}$ (above) and ${\psi=0}$ (below). This clearly illustrates the equivalence of the $\phi$ and $\psi$ coordinates in different orientations. In particular, two thin tori around ${\hat\rho=0}$ and around ${\hat\rho=1}$ do not intersect. }
\label{torus}
\end{figure}

The singularities at ${\hat\rho=0}$ and at ${\hat\rho=1}$ (i.e.~${\sqrt\Lambda\,\rho=0,\pi/2}$) are coordinate singularities (poles) since the Einstein universe is homogeneous and isotropic. Their meaning can be clearly seen in the familiar representation of a space of constant positive curvature as a three-dimensional hypersphere ${{x_1}^2+{x_2}^2+{x_3}^2+{x_4}^2=1}$ in a four-dimensional Euclidean space ${\d s^2={\d x_1}^2+{\d x_2}^2+{\d x_3}^2+{\d x_3}^2}$. Such representations of the spatial parts of the metrics (\ref{Estatmetric}) and (\ref{Estatmetric3}) are given by
 $$ \begin{array}{lcccc}
 \cos\chi &=&x_1&=& \sqrt{1-\hat\rho^2}\cos\psi, \\[2pt]
 \sin\chi\cos\theta &=&x_2&=& \sqrt{1-\hat\rho^2}\sin\psi, \\[2pt]
 \sin\chi\sin\theta\cos\phi &=&x_3&=& \hat\rho\cos\phi, \\[2pt]
 \sin\chi\sin\theta\sin\theta &=&x_4&=& \hat\rho\sin\phi. \\[2pt]
 \end{array} $$
 Consequently
 $$ \begin{array}{cc}
 {x_1}^2+{x_2}^2=1-\hat\rho^2, &\qquad {x_3}^2+{x_4}^2=\hat\rho^2, \\[5pt]
 {\displaystyle{x_2\over x_1}=\tan\psi}, &\qquad {\displaystyle{x_4\over x_3}=\tan\phi}.
 \end{array} $$
 It can thus be seen that the surfaces on which ${\hat\rho=}$~const. are a family of tori. They are surfaces spanned by~$\phi$ and~${\psi}$, which are both periodic. The angular character of these two coordinates is illustrated in figure~\ref{torus}.

The complete regular Einstein static space-time (i.e.~one without conical singularities) can thus be described by the metric~(\ref{Estatmetric5}) with ${t\in(-\infty,\infty)}$, ${\rho\in[0,\pi/2\sqrt\Lambda]}$, ${\phi\in[0,2\pi)}$ and ${z\in[0,2\pi/\sqrt\Lambda)}$, corresponding to ${\psi\in[0,2\pi)}$.

\section{The Linet--Tian solution}

The generalisation of the Levi-Civita solution to include a cosmological constant was obtained by Linet \cite{Linet86} and Tian \cite{Tian86}. It is usually expressed in the form
 \begin{equation}
 \begin{array}{l}
 \d s^2=Q^{2/3}\Big( -P^{-2(1-8\sigma+4\sigma^2)/3\Sigma}\,\d t^2 \\
 \hskip4.7pc
 +\,P^{-2(1+4\sigma-8\sigma^2)/3\Sigma}\,\d z^2 \\[2pt]
 \hskip4.7pc +\,C^{2}P^{4(1-2\sigma-2\sigma^2)/3\Sigma}\,\d\phi^2 \Big)
 +\d\rho^2,
 \end{array}
 \label{LinetTianmetric}
 \end{equation}
 where ${\Sigma=1-2\sigma+4\sigma^2}$, $\rho$ is a proper radial distance and
 \begin{equation}
 Q(\rho)={1\over\sqrt{3\Lambda}}\sin(\sqrt{3\Lambda}\,\rho), \qquad
 P(\rho)={2\over\sqrt{3\Lambda}}\tan\bigg({\sqrt{3\Lambda}\over2}\,\rho\bigg).
 \label{PQdefs}
 \end{equation}
 Note that ${\sqrt{-g}=CQ}$. This is, in fact, the unique static (apparently) cylindrically symmetric vacuum solution. It admits both cases ${\Lambda>0}$ and ${\Lambda<0}$ (in which trigonometric functions are replaced by hyperbolic ones).

It may immediately be seen that, both as ${\Lambda\to0}$ and as ${\rho\to0}$, the metric (\ref{LinetTianmetric}) approaches the form
 \begin{equation}
 \begin{array}{l}
 \d s^2 =-\rho^{4\sigma/\Sigma}\d t^2
 +\rho^{-4\sigma(1-2\sigma)/\Sigma}\d z^2 \\[2pt]
 \hskip6pc +C^{2}\rho^{2(1-2\sigma)/\Sigma}\d\phi^2 +\d\rho^2,
 \end{array}
 \label{LClimitmetric}
 \end{equation}
 which is the Levi-Civita metric. For ${0<\sigma<{1\over4}}$, this may be interpreted as a cylindrically symmetric space-time in which $\sigma$ represents the mass per unit length of a source along the axis ${\rho=0}$, and there is an additional conicity parameter~$C$. To retain the interpretation of this limit, it will be assumed below that the parameter $\sigma$ in the Linet--Tian metric is also restricted to the range ${\sigma\in[0,{1\over4}]}$.

When ${\sigma=0}$, both metrics (\ref{LinetTianmetric}) and (\ref{LClimitmetric}) approach Minkowski space near the axis at ${\rho=0}$, but with a conical singularity (or cosmic string) with deficit angle ${2\pi(1-C)}$ if ${\phi\in[0,2\pi)}$ and ${\phi=2\pi}$ is identified with ${\phi=0}$. Indeed, with ${t,z}$ constant, the circumference of a small circle divided by its radius is ${(\int_0^{2\pi}\!C\rho\,\d\phi)/\rho=2\pi C}$.

If ${\sigma\ne0}$, the limit as ${\rho\to0}$ corresponds to a curvature singularity for both metrics. However, the Linet--Tian metric (\ref{LinetTianmetric}) with ${\Lambda>0}$ has an additional curvature singularity at ${\rho=\pi/\sqrt{3\Lambda}}$, and the proper ``radial'' coordinate $\rho$ has the finite range ${(0,\pi/\sqrt{3\Lambda})}$. Thus, in addition to the presence of a source along the axis at ${\rho=0}$, there must exist another source near ${\rho=\pi/\sqrt{3\Lambda}}$. It follows that the metric~(\ref{LinetTianmetric}) can at most represent the field in a vacuum region with a cosmological constant {\em between two} concentric cylindrical-like sources.

Both the Linet--Tian metric (\ref{LinetTianmetric}) and the Levi-Civita metric (\ref{LClimitmetric}) are invariant under a transformation which replaces the parameter~$\sigma$ with ${1/4\sigma}$ and interchanges the coordinates $\phi$ and~$z$ (with a rescaling to reflect the change of conicity). However, this transformation takes $\sigma$ outside the range we wish to consider.

More importantly, the Linet--Tian metric (\ref{LinetTianmetric}) with ${\Lambda>0}$, ${C=1}$ is also invariant under the transformation\footnote{
  The equivalent transformation with ${\displaystyle \sigma=-{1-4\sigma'\over2(1+2\sigma')}}$ could alternatively be used, but this would include values of~$\sigma$ outside our assumed range. }
 \begin{eqnarray}
 \rho &=& {\pi\over\sqrt{3\Lambda}}-\rho' \nonumber\\
 t &=& \Big({4\over3\Lambda}\Big)^{(1-8\sigma+4{\sigma}^2)/3\Sigma}t', \nonumber\\
 \phi &=& \Big({4\over3\Lambda}\Big)^{-2(1-2\sigma-2{\sigma}^2)/3\Sigma}z', \label{invariance}\\
 z &=& \Big({4\over3\Lambda}\Big)^{(1+4\sigma-8{\sigma}^2)/3\Sigma}\phi', \nonumber\\
 \sigma &=& {1-4\sigma'\over4(1-\sigma')}. \nonumber
 \end{eqnarray}
 This demonstrates an equivalence between members of the family of Linet--Tian space-times with these different values of $\sigma$ and $\sigma'$. In this equivalence, both the character of the two curvature singularities, and the roles of the coordinates $\phi$ and $z$, are interchanged. Thus, if the Linet--Tian metric~(\ref{LinetTianmetric}) with ${\Lambda>0}$ and a particular value of~$\sigma$ is interpreted as having a source of strength~$\sigma$ near ${\rho=0}$ in the $z$-direction, then it could also be interpreted as having a similar source of strength~${\displaystyle \sigma'=\frac{1-4\sigma}{4(1-\sigma)}}$ near ${\rho=\pi/\sqrt{3\Lambda}}$ in the $\phi$-direction.

The Linet-Tian solution is usually assumed to have cylindrical symmetry. Consequently, the $z$-coordinate is normally taken to have the range ${(-\infty,\infty)}$, so that it can always be rescaled. However, in view of the interchange of $\phi$ and $z$ in the invariance (\ref{invariance}), it is clear that both of these coordinates have the same character. In particular, they are {\em both periodic}. This is consistent with the expectation that a static space-time with a positive cosmological constant would be spatially closed. Then, since neither of these coordinates can be arbitrarily rescaled, it follows that the general Linet--Tian solution with ${\Lambda>0}$ must contain {\em two} conicity parameters associated with possible deficit angles at both singularities.

In view of these properties, it is appropriate to relabel the coordinate $z$ as $\psi$ to reflect its character as an angle. Then, with two conicity parameters $B$ and $C$, we suggest that the Linet--Tian space-time with ${\Lambda>0}$ and ${\sigma\in[0,{1\over4}]}$ should preferably be given by the metric
 \begin{equation}
 \begin{array}{l}
 \d s^2=Q^{2/3}\Big( -P^{-2(1-8\sigma+4\sigma^2)/3\Sigma}\,\d t^2 \\[5pt]
 \hskip4pc +B^{2}P^{-2(1+4\sigma-8\sigma^2)/3\Sigma}\,\d\psi^2 \\[5pt]
 \hskip4pc +C^{2}P^{4(1-2\sigma-2\sigma^2)/3\Sigma}\,\d\phi^2 \Big)
 +\d\rho^2,
 \end{array}
 \label{LinetTianmetric2}
 \end{equation}
 where ${\Sigma=1-2\sigma+4\sigma^2}$, $Q(\rho)$ and $P(\rho)$ are given by (\ref{PQdefs}), and ${\phi,\psi\in[0,2\pi)}$. Apart from the cosmological constant, this metric has three parameters $B$, $C$ and $\sigma$.

It will be shown in the following section that the values of $B$ and $C$ can be established, for example, by matching this solution across a surfaces on which $\rho$ is a constant to a corresponding {\em toroidal} surface of the Einstein static universe.

For use below, it is also convenient to express the metric (\ref{LinetTianmetric2}) in the form
 \begin{equation}
 \begin{array}{l}
 \d s^2 ={\displaystyle \cos^{4/3}\!\Big(\!{\sqrt{{\textstyle{3\over4}}\Lambda}}\,\rho \Big)
 \Big( -P^{2p_0}\d t^2 } \\[8pt]
 \hskip4pc +B^{2}P^{2p_1}\d\psi^2
 +C^{2}P^{2p_2}\d\phi^2 \Big)+\d\rho^2,
 \end{array}
 \label{LinetTianmetric3}
 \end{equation}
 where ${P(\rho)}$ remains as given in (\ref{PQdefs}) and
 \begin{equation}
 p_0={2\sigma\over\Sigma}, \qquad p_1={2\sigma(2\sigma-1)\over\Sigma}, \qquad
  p_2={1-2\sigma\over\Sigma}.
  \label{parampsigma}
 \end{equation}
 In particular, it may be noticed that these constants satisfy the constraints
 $$ p_0+p_1+p_2=1, \qquad p_0^{\,2}+p_1^{\,2}+p_2^{\,2}=1. $$

\section{Matching conditions}

We will here consider removing a region around one of the two singularities of the Linet--Tian solution with ${\Lambda>0}$ and replacing it by an appropriate region of the Einstein static universe. Specifically, we will match the two metrics on a surface ${\rho=\rho_1}$. We may either take the Linet--Tian metric for ${\rho>\rho_1}$ and the Einstein metric for ${\rho<\rho_1}$, or vice-versa. In either case, the corresponding toroidal region of the Einstein static universe has a metric which can now be written in the form
 \begin{equation}
 \begin{array}{l}
\d s^2=-{\displaystyle A_1^{\,2}\,\d t^2 + {B_1^{\,2}\over\Lambda}\cos^2\Big(\sqrt\Lambda\,(\rho-\rho_{0})\Big)\,\d\psi^2} \\[7pt]
 \hskip4pc
 + {\displaystyle{C_1^{\,2}\over\Lambda}\sin^2\Big(\sqrt\Lambda\,(\rho-\rho_{0})\Big)\, \d\phi^2 + \d\rho^2},
 \end{array}
 \label{Estatmetric6}
 \end{equation}
 where ${\phi,\psi\in[0,2\pi)}$. This metric, which is a modification of (\ref{Estatmetric5}), includes four additional parameters $A_1$, $B_1$, $C_1$ and $\rho_{0}$. These correspond to a rescaling, allowances for possible deficit angles and a shift in the proper radial coordinate~$\rho$.

The required junction conditions are that the two metrics (\ref{LinetTianmetric2}) and (\ref{Estatmetric6}) and their first derivatives match across the surfaces ${\rho=\rho_1}$. These provide six conditions
 \begin{eqnarray}
 &&Q(\rho_1)\>P(\rho_1)^{-(1-8\sigma+4\sigma^2)/\Sigma} = A_1^{\,3}, \label{Acond}\\[5pt]
 &&Q(\rho_1)\>P(\rho_1)^{-(1+4\sigma-8\sigma^2)/\Sigma} \nonumber\\
 &&\hskip5pc=
 {B_1^{\,3}\over B^{\,3}\Lambda^{3/2}}\cos^3\Big(\sqrt\Lambda(\rho_1-\rho_{0})\Big), \label{Bcond} \\[5pt]
 &&Q(\rho_1)\>P(\rho_1)^{2(1-2\sigma-2\sigma^2)/\Sigma} \nonumber\\
 &&\hskip5pc=
 {C_1^{\,3}\over C^{\,3}\Lambda^{3/2}}\sin^3\Big(\sqrt\Lambda(\rho_1-\rho_{0})\Big), \label{Ccond} \\[5pt]
 &&{Q'(\rho_1)\over Q(\rho_1)} -\bigg({1-8\sigma+4\sigma^2\over1-2\sigma+4\sigma^2}\bigg){P'(\rho_1)\over P(\rho_1)} = 0, \label{Aprime} \\[5pt]
 &&{Q'(\rho_1)\over Q(\rho_1)} -\bigg({1+4\sigma-8\sigma^2\over1-2\sigma+4\sigma^2}\bigg){P'(\rho_1)\over P(\rho_1)} \nonumber\\
 &&\hskip5pc=
 -3\sqrt\Lambda\tan\Big(\sqrt\Lambda(\rho_1-\rho_{0})\Big), \label{Bprime} \\[5pt]
 &&{Q'(\rho_1)\over Q(\rho_1)} +2\bigg({1-2\sigma-2\sigma^2\over1-2\sigma+4\sigma^2}\bigg) {P'(\rho_1)\over P(\rho_1)} \nonumber\\
 &&\hskip5pc=
 3\sqrt\Lambda\cot\Big(\sqrt\Lambda(\rho_1-\rho_{0})\Big), \label{Cprime}
 \end{eqnarray}
 in which, from the definitions (\ref{PQdefs}),
 $$ {Q'(\rho_1)\over Q(\rho_1)}=\sqrt{3\Lambda}\cot(\sqrt{3\Lambda}\,\rho_1), \qquad {P'(\rho_1)\over P(\rho_1)} =
 {\sqrt{3\Lambda}\over\sin(\sqrt{3\Lambda}\,\rho_1)}. $$

The first three equations (\ref{Acond})--(\ref{Ccond}) effectively determine the constants $A_1$, $B_1/B$ and $C_1/C$ in terms of the parameters $\Lambda$, $\sigma$, and $\rho_1$.

For any solution with given values for $\Lambda$ and~$\sigma$, the condition (\ref{Aprime}) becomes
 \begin{equation}
 \cos(\sqrt{3\Lambda}\,\rho_1) =
 {1-8\sigma+4\sigma^2\over1-2\sigma+4\sigma^2}.
 \label{cond2}
 \end{equation}
 Notice that the right hand side of this expression decreases monotonically from 1 to $-1$ as $\sigma$ increases from 0 to ${1\over4}$, so it determines a {\em unique} value for $\rho_1$ explicitly. With this value, it follows that
   $$ \sin(\sqrt{3\Lambda}\,\rho_1) =
 {2\sqrt3\sqrt\sigma\sqrt{1-4\sigma}\sqrt{1-\sigma}\over1-2\sigma+4\sigma^2}. $$
 Hence
 \begin{equation}
 \begin{array}{l}
 {\displaystyle Q(\rho_1)={2\sqrt\sigma\sqrt{1-4\sigma}\sqrt{1-\sigma}\over\sqrt{\Lambda}\,(1-2\sigma+4\sigma^2)}, } \\[8pt]
 {\displaystyle P(\rho_1)={2\sqrt\sigma\over \sqrt\Lambda \sqrt {1-4\sigma} \sqrt{1-\sigma}},}
 \end{array}
 \end{equation}
 and
 \begin{equation}
 \begin{array}{l}
 {\displaystyle {Q'(\rho_1)\over Q(\rho_1)} = {\sqrt{\Lambda}\,(1-8\sigma+4\sigma^2)
 \over 2\sqrt\sigma\sqrt{1-4\sigma}\sqrt{1-\sigma}}, } \\[12pt]
 {\displaystyle {P'(\rho_1)\over P(\rho_1)} = {\sqrt{\Lambda}\,(1-2\sigma+4\sigma^2)
 \over 2\sqrt\sigma\sqrt{1-4\sigma}\sqrt{1-\sigma}}.}
 \end{array}
 \end{equation}
 With these expressions, equations (\ref{Bprime}) and (\ref{Cprime}) are in fact identical. Each implies that
 \begin{equation}
 \tan^2\Big(\sqrt\Lambda\,(\rho_1-\rho_0)\Big) ={4\sigma(1-\sigma)\over1-4\sigma}.
 \label{sigmaresult}
 \end{equation}
 For $\rho_1$ in the given range, this defines a unique (positive) value for the parameter~$\rho_0$.

\goodbreak
We must now investigate the (relative) values of the constants $A_1$, $B_1$, $C_1$ and the conicity parameters $B$ and~$C$. These are given by the equations (\ref{Acond})--(\ref{Ccond}) in the form
 \begin{eqnarray}
 A_1 &=& Q(\rho_1)^{1/3}\>P(\rho_1)^{-(1-8\sigma+4\sigma^2)/3\Sigma} , \\[5pt]
 B_1 &=& B\,\sqrt\Lambda\, {Q(\rho_1)^{1/3}\>P(\rho_1)^{-(1+4\sigma-8\sigma^2)/3\Sigma} \over \cos\Big(\sqrt\Lambda\,(\rho_1-\rho_{0})\Big)}, \label{Bcond2} \\[5pt]
 C_1 &=& C\,\sqrt\Lambda\, {Q(\rho_1)^{1/3}\>P(\rho_1)^{2(1-2\sigma-2\sigma^2)/3\Sigma} \over \sin\Big(\sqrt\Lambda\,(\rho_1-\rho_{0})\Big)}. \label{Ccond2}
 \end{eqnarray}
 The first of these determines the value of $A_1$ explicitly. The other two need to be interpreted more carefully.

Consider first the case in which the region around the curvature singularity of~\eqref{LinetTianmetric2} at ${\rho=0}$ is replaced by a corresponding toroidal region of the Einstein static universe, so that the space-time is described by the metric (\ref{Estatmetric6}) with ${\rho\in[\rho_0,\rho_1)}$, and the metric (\ref{LinetTianmetric2}) with ${\rho\in[\rho_1,\pi/\sqrt{3\Lambda})}$. In this case, the pole at ${\rho=\rho_0}$ is regular (with no conical singularity) if ${C_1=1}$. The corresponding value of $C$ can then be determined from (\ref{Ccond2}) explicitly. Any other value of $C$ would lead to a conical singularity on the axis of the toroidal region of the Einstein static universe as determined by the value of~$C_1$ given by~(\ref{Ccond2}).

Interestingly, the total mass of the dust within this toroidal section of the Einstein universe is
 $$ \int_{\rho_0}^{\rho_1}\int_0^{2\pi}\int_0^{2\pi}\mu\,\sqrt{g_3}\,\d\rho\,\d\psi\,\d\phi
 ={2\pi B_1C_1\over\sqrt\Lambda}\,{\sigma(1-\sigma)\over(1-4\sigma^2)}. $$
 Since the length of this toroid is approximately ${2\pi B_1/\sqrt\Lambda}$ near ${\rho=\rho_0}$, and ${C_1=1}$, the mass per unit length of the toroid is ${\sigma(1-\sigma)/(1-4\sigma^2)}$, which is approximately $\sigma$ for small values. This is consistent with the result that is expected when ${\Lambda\to0}$. Notice, however, that this is not the sole ``source'' for this space-time. The contribution from the remaining singularity must also be taken into account.

In the opposite case in which the other region, namely that around the curvature singularity at ${\rho=\pi/\sqrt{3\Lambda}}$, is replaced by a corresponding toroidal region of the Einstein static universe, the space-time is described by the metric (\ref{LinetTianmetric2}) with ${\rho\in(0,\rho_1]}$, and the metric (\ref{Estatmetric6}) with ${\rho\in(\rho_1,\rho_0+\pi/2\sqrt{\Lambda}]}$. In this case, the pole at ${\rho=\rho_0+\pi/2\sqrt{\Lambda}}$ is regular if ${B_1=1}$. The value of $B$ can then be determined from (\ref{Bcond2}). Any other value of $B$ would lead to a conical singularity in the toroidal region of the Einstein static universe (a closed cosmic string).

\section{The limit as ${\sigma\to0}$}

For the case in which $\sigma$ vanishes, the Linet--Tian metric (\ref{LinetTianmetric2}) with ${\Lambda>0}$ becomes
 \begin{equation}
 \begin{array}{l}
 \d s^2={\displaystyle \cos^{4/3}\bigg({\sqrt{3\Lambda}\over2}\rho\bigg)\,(-\d t^2
 +B^{2}\,\d\psi^2) +\d\rho^2 } \\[10pt]
 \hskip2pc {\displaystyle +{4C^2\over{3\Lambda}}
 \sin^2\bigg({\sqrt{3\Lambda}\over2}\rho\bigg)
 \cos^{-2/3}\bigg({\sqrt{3\Lambda}\over2}\rho\bigg)\,\d\phi^2, }
 \end{array}
 \label{}
 \end{equation}
 Putting $p=\cos^{2/3}(\sqrt{3\Lambda}\,\rho/2)$ gives
 \begin{equation}
 \begin{array}{l}
 \d s^2 =p^2(-\d t^2+B^{2}\,\d\psi^2) \\[5pt]
 \hskip2pc +{\displaystyle {4C^2\over3\Lambda}{(1-p^3)\over p}\,\d\phi^2
 +{3\over\Lambda}{p\over(1-p^3)}\,\d p^2, }
 \end{array}
 \label{BadSmetric3}
 \end{equation}
 where ${p\in[0,1]}$ with ${p=1}$ representing the axis ${\rho=0}$. (The Linet--Tian metric with ${\Lambda<0}$ also reduces to this form, but the range of~$p$ is altered to retain a Lorentzian metric.)

The metric (\ref{BadSmetric3}) is known to be of type~D, and clearly belongs to the family of non-expanding Pleba\'nski--Demia\'nski solutions whose general form is given in equation (16.27) of \cite{GriPod09}. For the particular case in which the parameters $\alpha$ and $\gamma$ both vanish, these solutions are given by the metric
 \begin{equation}
 \d s^2=
 p^2\Big(-{\cal Q}\,\d t^2 +{1\over{\cal Q}}\,\d q^2 \Big)  +{{\cal P}\over p^2}\,\d\varphi^2  +{p^2\over{\cal P}}\,\d p^2  \,,
 \label{nonExpMetric}
 \end{equation}
 where
 $$ {\cal Q}(q)= \epsilon_0+\epsilon\,q^2, \qquad
 {\cal P}(p)= 2n\,p -\epsilon\,p^2 -{\textstyle{1\over3}}\Lambda\,p^4. $$
 Clearly the metric (\ref{BadSmetric3}) is a particular member of this family for which ${\epsilon_0=1}$, ${\epsilon=0}$, ${n={1\over6}\Lambda}$ and the coordinates $\phi$ and $\psi$ have been rescaled. This is thus a generalisation of the BIII metric which includes a cosmological constant. It is also a particular type D solution of Kundt's class. However, none of these solutions are well understood physically.

\section{Extension to higher dimensions}

Interestingly, the solution described above can be extended to any higher number ($D$) of dimensions. The metric in this case is given by
 \begin{equation}
 \d s^2=R(\rho)^\alpha\bigg(\! -S(\rho)^{2p_0}\,\d t^2 +\sum_{i=1}^{D-2}C_i^{\,2}S(\rho)^{2p_i}\,\d\phi_i^{\,2} \!\bigg) +\d\rho^2,
 \label{LThigherD}
 \end{equation}
where ${\phi_i\in[0,2\pi)}$ are angular coordinates,
$$ R(\rho)=\cos(\beta\rho), \qquad S(\rho)=\tan(\beta\rho), $$
with
$$ \alpha={4\over D-1}, \qquad \beta=\sqrt{{(D-1)\over2(D-2)}\Lambda}, $$
the constants $C_i$ are corresponding conicity parameters and the constants $p_i$ satisfy the constraints
$$ \sum_{i=0}^{D-2}\,p_i=1, \qquad \sum_{i=0}^{D-2}\,p_i^{\,2}=1. $$
Such a metric satisfies Einstein's vacuum field equations ${R_{\mu\nu}=\frac{2}{D-2} \Lambda\, g_{\mu\nu}}$ with a cosmological constant ${\Lambda>0}$.

The metric \eqref{LThigherD} clearly reduces to the Linet--Tian solution in the form \eqref{LinetTianmetric3} with \eqref{PQdefs} when ${D=4}$, in which case the constants $p_i$ are expressed in terms of a single parameter~$\sigma$ using \eqref{parampsigma}.

Notice that the analogous extension of the Linet--Tian solution to higher dimensions for the case when ${\Lambda<0}$ has been recently given in \cite{SarTek09}.

There exists an important special subcase of the metric \eqref{LThigherD} in which ${p_0=0}$, ${p_i=0}$ for ${i=1,\dots, D-3}$ and ${p_{_{D-2}}=1}$. In view of the relations \eqref{parampsigma}, this corresponds to the ``source-free'' limit ${\sigma=0}$ in the Linet--Tian metric, as described in the previous section for the case ${D=4}$. This particular metric reads
 \begin{eqnarray}
 && \hskip-4pc \d s^2=\cos^\alpha(\beta\rho)\Big(\! -\d t^2 +C^{2}\tan^{2}(\beta\rho)\,\d\phi^{2} \nonumber\\
 && \hskip3pc +\sum_{i=1}^{D-3}C_i^{\,2}\,\d\phi_i^{\,2} \Big)+\d\rho^2,
 \label{LThigherDspecial}
 \end{eqnarray}
where we have relabeled the coordinate ${\phi_{_{D-2}}}$ as ${\phi}$.

The ${\Lambda<0}$ counterpart of the solution \eqref{LThigherDspecial} has been identified in \cite{SarTek09b} as a $D$-dimensional generalisation of a ``second anti-de Sitter universe'' of \cite{Bonnor08} and as a special case of an ``AdS soliton'' of \cite{HorMey99}.

\section{Conclusions}

It is already well-known that the vacuum Linet--Tian solution with ${\Lambda>0}$ has two curvature singularities at ${\rho=0}$ and ${\rho=\pi/\sqrt{3\Lambda}}$. It has been argued here that this space-time is essentially toroidally symmetric with the singularities located at the poles, which form closed axes. The familiar coordinates $\phi$ and $z$ have been shown to be both periodic, so that neither can be rescaled and {\em two} corresponding conicity parameters should, therefore, be included in the metric. In view of these properties, it is suggested that the solution should be given in the form of the metric~(\ref{LinetTianmetric2}).

The two curvature singularities are generally different. However, they have a similar character and are related by the invariance property~(\ref{invariance}).

It has been shown that a region of the vacuum Linet--Tian solution can be matched across a toroidal surface to part of the static Einstein universe containing dust. This clarifies the above interpretation. However, since the equation (\ref{cond2}) has just a single solution for $\rho_1$ for any given value of~$\sigma$, it is not possible in this way to replace both curvature singularities simultaneously with toroidal sources having a finite vacuum region between them.

In the limit as ${\Lambda\to0}$, this solution reduces to the well-known Levi-Civita solution in which the parameter~$\sigma$ may be interpreted as a mass per unit length of a cylindrical source. Interestingly, such an interpretation also applies to the parameter~$\sigma$ in the Linet--Tian solution with ${\Lambda>0}$ when considering the source around ${\rho=0}$.

We also presented a higher-dimensional generalisation of the Linet--Tian vacuum solution in the case when ${\Lambda>0}$. This forms a natural counterpart of the recently discussed family of metrics with a negative cosmological constant which involves a ``second anti-de Sitter universe''  and a special case of an ``AdS soliton''.

\section*{Acknowledgements}

The authors are grateful to Martin \v{Z}ofka for some helpful comments. This work was supported by the grant GA\v{C}R~202/08/0187 and by the project LC06014 of the Czech Ministry of Education.

\end{document}